\documentclass{emulateapj}
\usepackage{epsfig}

\newcommand{\hi}{H{\sc i~}}

\shorttitle{Abundant dust found in intergalactic space.}
\shortauthors{Xilouris et al.}

\begin{document}

\title{Abundant dust found in intergalactic space.}

\author{E. Xilouris}
\affil{Institute of Astronomy \& Astrophysics, National Observatory of Athens, I. Metaxa \& V. Pavlou, P. Penteli, GR-15236, Athens, Greece}
\email{xilouris@astro.noa.gr}

\author{P. Alton}
\affil{Environmental Modelling and Earth Observation group, Department of Geography, University of Wales Swansea, Singleton Park, Swansea SA2 8PP, UK}
%\email{P.Alton@swansea.ac.uk}

\author{J. Alikakos\altaffilmark{1}}
\affil{Institute of Astronomy \& Astrophysics, National Observatory of Athens, I. Metaxa \& V. Pavlou, P. Penteli, GR-15236,
Athens, Greece}
%\email{johnal@astro.noa.gr}

\author{K. Xilouris}
\affil{University of Arizona, Steward Observatory, 933 N. Cherry Av. Bldg 65, Tucson, AZ 85721, USA}
%\email{kxiluri@as.arizona.edu}

\author{P. Boumis}
\affil{Institute of Astronomy \& Astrophysics, National Observatory of Athens, I. Metaxa \& V. Pavlou, P. Penteli, GR-15236, Athens, Greece}
%\email{ptb@astro.noa.gr}

\and 

\author{C. Goudis\altaffilmark{1}}
\affil{Institute of Astronomy \& Astrophysics, National Observatory of Athens, I. Metaxa \& V. Pavlou, P. Penteli, GR-15236, Athens, Greece}
%\email{cgoudis@astro.noa.gr}

\altaffiltext{1}{Department of Physics, University of Patras, GR-26500, Rio-Patras, Greece}
\clearpage
\begin{abstract}
Galactic dust constitutes approximately half of the elements more massive than helium produced in stellar
nucleosynthesis.
Notwithstanding the formation of dust grains in the dense, cool atmospheres of late-type stars, there still remain huge  uncertainties
concerning the origin and fate of galactic stardust.
In this paper, we identify the intergalactic medium (i.e. the region between gravitationally-bound galaxies) as a major sink for galactic
dust. We discover a systematic shift in the colour of background galaxies viewed through the intergalactic medium of the nearby
 M81 group. This reddening coincides with atomic, neutral gas previously detected between the group members.
The dust-to-\hi mass ratio is high (1/20) compared to that of the solar neighborhood (1/120)
suggesting that the dust originates from the centre of one or more of the galaxies in the group.
Indeed, M82, which is known to be ejecting dust and gas in a starburst-driven superwind, is cited as the probable main source.
\end{abstract}

\clearpage
%\clearpage

\section{Introduction}

The mass contained within galactic dust is only 1\% that contained within interstellar gas and 
amounts to less than 0.1\% of the total mass of the galaxy. Nevertheless, galactic dust has a
major influence on both our perception of galactic structure (through extinction) and the processes taking place in interstellar medium.
Indeed, dust grains provide the reaction sites necessary for the formation of complex  molecules (Herbst 2001) and constitute the building blocks of
planet formation (Greaves et al. 2004).

The balance between the formation and the removal of dust in spiral galaxies, like our own, is far from clear.
Cool, dense atmospheres of stars that have left the main sequence (red giants) are currently identified as a primary source of dust
in our own Galaxy (0.04 M$_{\odot}$/year; Whittet 1992). However, the destruction 
rate of dust grains from interstellar shocks is predicted to be
relatively high suggesting other primary dust sources are yet to be identified (Seab 1988).

Intense star-formation in galactic disks is known to eject dust several kiloparsecs out of 
the main stellar plane but, thus far, it is not clear
whether such material escapes entirely from the galaxy or returns to the main disk in a 
galactic-scale, convective, mixing process (Heckman,  Armus, \& Miley  1990; Howk \& Savage 1997).
In this paper, we examine the photometric colour of background galaxies viewed through the intergalactic medium of the M81 group. Any systematic
reddening of the background objects, compared to the same population viewed adjacent to the group, reveals the presence of dust
residing between the group members (galactic dust absorbs more blue than red light thus both attenuating and reddening light
from background sources). The M81 group is a suitable testing ground for theories of dust loss into the intergalactic medium for several reasons.
Firstly, it constitutes one of the nearest examples of a tidally interacting ensemble of galaxies (distance 3.6 Mpc; Freedman et al. 1994).
Secondly, copious amounts of neutral atomic hydrogen gas have been found connecting the various group members in a series of bridges and
filaments (Yun, Ho, \& Lo 1994). Thirdly, one of the three main galaxies in the group (M82) 
is known to be expelling dust and neutral gas away from its
starburst core towards the intercluster medium (Ichikawa et al. 1994; Alton, Davies, \& Bianchi  1999; Engelbracht et al. 2006).

Comparing the mean colour of background objects is well established as a technique for probing dust outside the main 
stellar disk (Zaritsky 1994; Lequeux, Dantel-Fort, \& Fort 1995; Boyle, Fong, \& Shanks 1988; Holwerda et al. 2005). 
However, the advent of large, optical imaging arrays, which incorporate several
contiguous CCDs, permit sufficient numbers of distant galaxies to be detected that we can begin to trace the distribution of
foreground dust rather than simply its mean optical depth.

\section{Observations and data analysis}

%%%%%%%%%%%%%%%%%%%%% FIG 1 %%%%%%%%%%%%%%%%%%%%%%%%%%%
\begin{figure}
\epsfig{figure=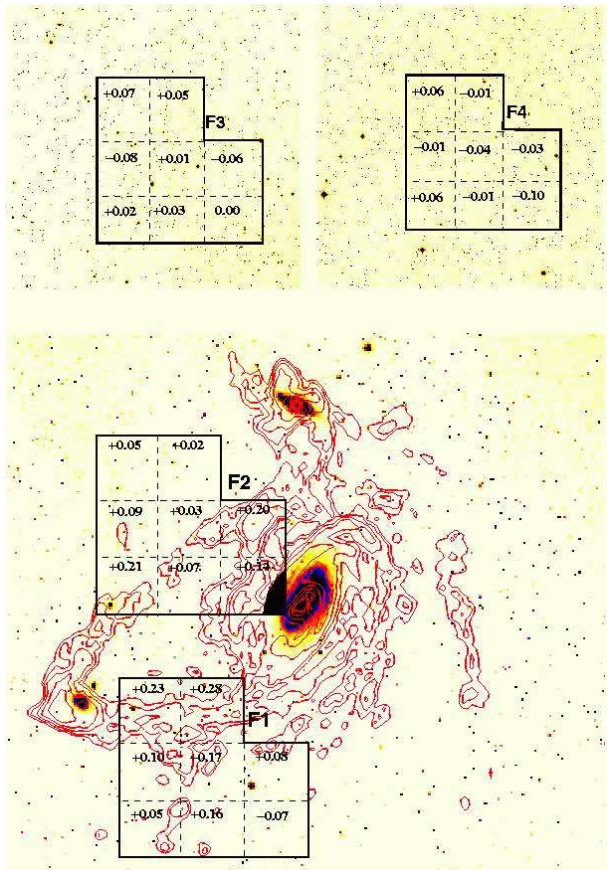, width=8truecm, angle=0}
\caption{Map of the M81 Group ($1.8 \times 1.8$ degrees) showing the atomic, neutral
hydrogen detected by Yun, Ho, \& Lo (1994) as contours ($3 \times 10^{19}$ cm$^{-2} \times 2^n$) and a B-band
image as greyscale (from the Digitized Sky Survey). Note that extensive, diffuse gas is present outside the optical disks. The position of the WFC is outlined as a solid line. F3 and F4
constitute control fields whilst F1 and F2 have been chosen to coincide with significant diffuse gas (note that the shaded part of F2 in the bottom right corner is dominated by emission coming
from the main body of M81 and was therefore masked and neglected from the further analysis).
Background objects detected in each field are divided
into 8 sub-fields and the B-I excess with respect to the colour of the control population are indicated in each box.}
\end{figure}

%%%%%%%%%%%%%%%%%%%%% FIG 2 %%%%%%%%%%%%%%%%%%%%%%%%%%%
\begin{figure*}
\begin{center}
\epsfig{figure=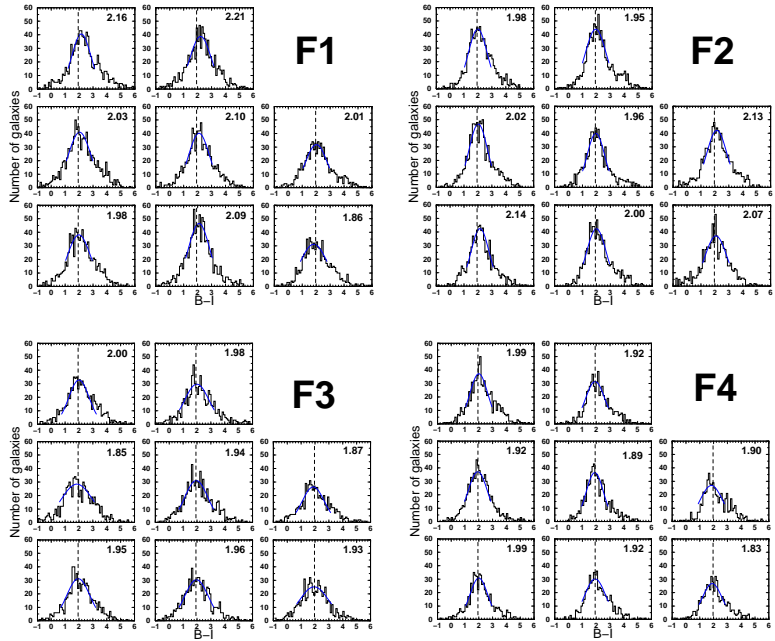, width=10.2truecm, angle=0}
\end{center}
\caption{
B-I histogram of the background galaxies detected in each of the 4 fields. A gaussian has been fitted
to the peak of the distribution in each sub-field in order to estimate the mode average (the latter is
shown top-right in each box).}
\end{figure*}

The Wide-Field Camera (WFC) mounted on the 2.5-m Isaac Newton Telescope (INT), 
La Palma, consists of four $2048 \times 4100$ pixel EEV CCDs (0.33''/pixel) and
provides a field-of-view of 0.29 deg$^{2}$. 
During four dark nights at the INT (March 2004), we have used the WFC
to observe the M81 group in the Johnson B and the Sloan I filters (wavelengths of 436 and 767 nm, respectively).
Of a total of four fields observed, two `on-source' regions (F1 and F2) correspond to
parts of the group where extended \hi emission has been detected by Yun, Ho, \& Lo (1994). The remaining two fields were situated
approximately 4 deg from the centre of the group and constituted control fields (F3 and F4). Figure 1 shows the position of our chosen
fields and Table 1 provides a summary of the observations.

The debiasing, flat-fielding and trimming
of the frames obtained with the WFC were carried out using standard IRAF and MIDAS image software packages. Fringes in the I-band, at a level of $\simeq$ 4\%, were removed by observing blank sky fields containing relatively few sources on each night of observation and for multiple positions in the sky. During each night, 3 or 4
standard star fields (Landoldt 1992) were observed in different airmasses and the measurements fitted with a
3-term calibration equation (photometric zero-point, airmass, color).
In this way, photometric accuracy, on any given night, is estimated to be 0.02 mag and 0.01 mag in the B and I bands, respectively.
Our limiting magnitudes in B and I are 27 and 24, respectively.

The photometrically-calibrated WFC images were processed using the SExtractor software (Bertin \& Arnouts 1996) 
which allows resolved objects of comparatively low
surface brightness (galaxies) to be separated from point sources of high surface brightness (foreground Galactic stars). Cosmic rays are
eliminated by imposing a minimum radius of 4 pixels or 1.32$''$ for a bonafide detection (threshold at which surface brightness decreases to
$50\%$ of its maximum value). We have followed previous authors in using SExtractor with the CLASS\_STAR parameter set to 0.92 but,
for all fields, the successful separation of galaxies from stars was assured by visual inspection by plotting magnitude against half-light radius
(stars appear as a well-defined locus in such a plot). The galaxy counts in our fields (objects per square degree per magnitude)
were in excellent agreement with those previous published in deep surveys (Metcalfe et al. 2001).
Typically, $\sim 5000$ galaxies were detected in both bands (B and I) for each of the four fields.

After galaxy-star separation the next step was to determine the  B-I colour of the background objects.
An important preliminary in this process was the subtraction of foreground reddening associated with dust in the Milky Way. This was achieved 
by referring to maps of Galactic B-V colour excess [E(B-V)] along lines of site to each of our fields (Schlegel, Finkbeiner, \& Davis 1998)
and applying a Galactic reddening law to convert E(B-V) to B-I
colour excess [E(B-I)=2.23$\times$E(B-V)]. For each field, foreground E(B-V) was sampled for 8 constituent sub-fields ($12' \times  12'$)
and the mean value used to correct the B-I colours for background objects detected in that particular field. The corrections
in B-I were -0.163, -0.120, -0.234 and -0.116 for F1, F2, F3 and F4, respectively, with a typical dispersion of 0.038 (root mean square).
Along with the photometric error (0.026 mag) of the Galactic E(B-V) maps (Schlegel, Finkbeiner, \& Davis 1998) the total uncertainty for the subtruction of the foreground Galactic
reddening becomes 0.046 mag.

\section{Results}

In Figure 2 we show the histograms of the B-I colour for the background
galaxies detected in each sub-field.
A smooth gaussian function was fitted to the peak of each histogram to yield average B-I colour.
The typical uncertainty of the peak value of the gaussian, as determined
by the fitting procedure, is 0.02 mag.
For the control fields, F3 and F4,  mean B-I was recorded as
1.935$\pm$0.048 and 1.920$\pm$0.049, respectively, indicating a B-I colour for the background population of 1.928$\pm$0.035. 
This colour is consistent with that found in deep imaging surveys (Driver et al. 1994). 

For the `on-source'
sub-fields, comprising F1 and F2, values in excess of 1.928 indicate reddening by dust residing in the intergalactic medium.
Taking into account both the uncertainty of the B-I colour for the background population 
(0.035 mag) and the uncertainty of the average B-I colour of the `on-source' fields (0.02 mag),
the uncertainty in the B-I colour excess of the galaxies in our fields become 0.04 mag.
Figure 2 reveals 10 sub-fields in F1 and F2 where such an excess is statistically significant (shift $\ge$ 0.08 mag i.e. $\ge2\sigma$). Notably, the greatest reddening
coincides with the densest concentrations of atomic hydrogen in F1 and F2 and thus indicates a V-band extinction (A$_{V}$) as high as
0.39 mag (using a total to selective extinction R$_{V} = 3.08$; Pei 1992). The excesses with respect to 
the control fields are indicated in Figure 1 (the number shown in each sub-field).

%%%%%%%%%%%%%%%%%%%%% FIG 3 %%%%%%%%%%%%%%%%%%%%%%%%%%%
\begin{figure}
\epsfig{figure=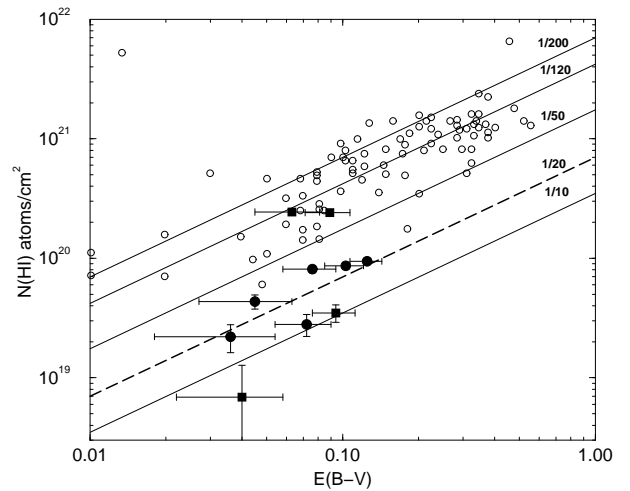, width=8truecm, angle=0}
\caption{
Surface density of atomic hydrogen [N(H{\sc i})] against colour excess [E(B-V)]. Open circles correspond to measurements for the Milky Way close to the solar circle (Bohlin, Savage, \& Drake 1978). These are compared with the corresponding values that we record for the on-source
fields F1 and F2 (solid circles and solid squares,
respectively) where significant reddening has been detected. Parallel lines indicate different
dust-to-\hi mass ratios.
The two uppermost points for F2 are known
to correspond to areas of low metallicity (Makarova et al. 2002).}
\end{figure}

The mass ratio of dust-to-gas is known to correlate with heavy metal abundance, at least for the Milky Way and other members
of the Local Group (Issa, MacLaren, \& Wolfendale 1990). Thus, for sub-fields where significant reddening is detected, we have 
plotted the dust-to-\hi mass ratio and
compared with the corresponding value in the solar neighborhood (Figure 3). 
The parallel lines in this plot indicate different dust-to-\hi mass ratios
assuming a value of 1/120 for the solar
neighborhood (Sodroski et al. 1997; see the figure caption for more
information on the symbols). 
The uncertainty in the surface density of the 
atomic hydrogen is derived by using a 0.8 mJy/beam r.m.s. as indicated in Yun, Ho, 
\& Lo (1994), while the uncertainty in E(B-V) is 0.018
[note that we have adopted a Galactic extinction law to convert from
B-I to B-V excess E(B-I)=2.23 $\times$ E(B-V)]. 
The intergalactic medium of the M81 group 
appears to be dust rich with
an average dust-to-\hi mass ratio of 1/20 with values typically ranging between
(1/10) and (1/50) [note that the two uppermost points for F2 in Figure 3
are known to correspond to areas of low metallicity (Makarova et al. 2002)]. 
This value is approximately 6 times  higher than that of the solar neighborhood 
(1/120 - Sodroski et al. 1997).

\section{Discussion and conclusions}

The implication of our finding (relatively high dust-to-\hi mass ratio in the 
intergalactic space) is that the intergalactic material originates from a metal-rich environment
since the outskirts of spiral galaxies such as M81 are epitomised by dust-to-gas mass ratios of less than 
the solar level (Cuillandre et al. 2001; Lequeux, Dantel-Fort, \& Fort 1995)
and metallicities several times lower than the solar abundance (Vila-Costas \& Edmunds 1992).
Indeed, at the D$_{25}$ radius ($12'$) of M81 (the optical `edge' of the stellar disk), the metallicity
has been measured as 1/4 solar (Vila-Costas \& Edmunds 1992) implying that if the intercluster material had been 
drawn out from the peripheries of
M81 it would have a dust-to-gas ratio of 4 times {\em less} than solar rather than our estimate of 6 
times {\em higher}. Equally, we can show
that the reddening we have detected cannot be attributed simply to the interstellar dust of M81 extending to extraordinarily large radii. Assuming a maximum 
V-band optical depth of 5 for the centre of M81 (e.g. Alton et al. 2001) and a maximum radial, exponential scalelength for interstellar dust of 5 kpc
(e.g. Xilouris et al. 1999) , the V-band optical depth in the sub-fields of F1 closest to M81 ($\sim$ 25 kpc) should not exceed 5$\times$exp(-25/5)=0.034. The corresponding
V-band extinction (0.037 mag) is nearly an order of magnitude lower than the attenuation we record for this region (A$_{V}\simeq$ 0.24 mag).

Given the relative abundance of the intergalactic dust with respect to the hydrogen gas we postulate the centre of M82 as a likely source for the dust.
An energetic, starburst-driven outflow is known to be entraining up to $10^{7}$ M$_{\odot}$ of dust away from the stellar disk towards the intergalactic
medium (Alton, Davies, \& Bianchi 1999). Simulations of the tidal interaction within the M81 group indicate
that any material ejected by M82 is likely to be well dispersed across the intergalactic medium in the current 
configuration (Yun 1999). The total \hi
gas residing outside the galactic disks of the group is $9.6\times 10^{8}$ M$_{\odot}$ (Yun, Ho, \& Lo 1994). 
Thus for a dust-to-\hi mass ratio of 1/20 (Figure 3) we infer a total dust mass
for the intergalactic medium of $4.8\times 10^{7}$ M$_{\odot}$. This is equivalent to the total amount of dust contained within a typical spiral galaxy such as M81 (Stickel et al. 2000).
Assuming that M82 constitutes the remnants of a spiral galaxy which has been violently disturbed by its tidal encounter with the other group members
(its current irregular morphology bears testament to this violent collision), the ensuing period of starburst-driven outflow appears to have been
a highly effective mechanism for removing grains from the stellar disk.

Although, currently, starburst activity appears to occur in only 1 in 10 galaxies, the phenomenon is believed to be far more common
at earlier epochs in the lifetime of the Universe (Heckman, Armus, \& Miley 1990). 
Our results imply, therefore, that the intergalactic medium is a significant sink for
dust grains originating in galactic disks. 
This dust together with the vast amounts of gas residing in these areas can be
used for the formation of tidal dwarfs, at least in areas close to the galaxtic disks (see
Makarova et al. 2002). The grain properties (i.e. sizes and composition) as well
as kinematical information on the dust grains and the gas are important parameters 
to study the fate of the material present in the intergalactic space between the
galaxies of the M81 group.
Several workers detect a systematic deficit in the density of quasars along lines of sight to
clusters of intermediate redshifts (z $\simeq$ 0.1) consistent with extinction levels close to that found in the 
present study (Boyle, Fong, \& Shanks  1988).
This implies that the intergalactic nature of the copious dust we have detected in the M81 group is not unusual.

\acknowledgments
We thank P. Papadopoulos, N. Kylafis, V. Charmandaris and P.-A. Duc for stimulating
discussions. We are also indebted to the anonymous referee for useful
comments and suggestions. The INT is operated on the island of La Palma by the
Isaac Newton Group 
in the Spanish Observatorio del Roque de los Muchachos of the Instituto de Astrofisica de 
Canarias. Observations at the INT were supported by the OPTICON access programme.
OPTICON has received research funding from the European Community's Sixth Framework Programme 
under contract number RII3-CT-001566

%\clearpage
%%%%%%%%%%%%%%%%%%%%% TABLE 1 %%%%%%%%%%%%%%%%%%%%
\begin{table*}
\begin{center}
\caption{A summary of the observations. \label{tbl-1}}
\begin{tabular}{lccccc}
\tableline\tableline
      &    R.A.    &    Decl.    &    Observation   &
\underline{~~Exposure Time~~}& Seeing \\
Field & (J2000.0)  &  (J2000.0)  &       Date       &
(B-band)~(I-band) &  ('')    \\
\hline
F1    & 10 00 00.5 & 68 32 36.5  & 2004 Mar 18    & 160 ~~~~~~ 85  & 1.3\\
F2    & 10 00 35.1 & 68 19 39.8  & 2004 Mar 20    & 165 ~~~~~~ 60  & 1.2\\
      &            &             & 2004 Mar 21    & 20  ~~~~~~~ 35 & 1.6\\
F3    & 10 30 55.1 & 72 14 35.9  & 2004 Mar 19    & 80  ~~~~~~~ 30 & 1.4\\
      &            &             & 2004 Mar 20    & 100 ~~~~~~ 55  & 1.2\\
F4    & 09 25 46.6 & 72 17 02.9  & 2004 Mar 21    & 180 ~~~~~~ 90  & 1.6\\
\hline
\end{tabular}
\tablecomments{Units of rihgt ascension are hours, minutes, and seconds,
units of declination are degrees,
arcminutes, and arcseconds. Exposoure time is given in minutes.}
\end{center}
\end{table*}
%%%%%%%%%%%%%%%%%%%%%%%%%%%%%%%%%%%%%%%%%%%%%%%%%%%%%%%%%%%%%%%%%%%%%%

\end{document}